\documentclass[aps,floatfix,prd,twocolumn,a4paper,10pt,citeautoscript,
preprintnumbers,superscriptaddress]{revtex4-1} 

\usepackage[english]{babel}	%%% LaTeX for English

\usepackage{amsmath}		%%% Mathematics AMS package 
\usepackage{amssymb}		%%% More math symbols
\usepackage{bm}			%%% Bold math font 
\usepackage{bbm}			%%% Font for ensembes

\usepackage{graphicx}	%%% Include graphics
\usepackage{graphics}	%%% Include graphics
\DeclareGraphicsExtensions{.jpg}	%%% Extensions for pdflatex

\usepackage{array}		%%% Additional tabular tools
\usepackage[colorlinks=true, pdfstartview=FitV,
			linkcolor=blue, citecolor=blue, urlcolor=blue]{hyperref}

\newcommand{\Equation}[2]{  \begin{equation}\label{#1}#2\end{equation} }
\newcommand{\Align}[2]{\begin{align}\label{#1}#2\end{align}}
\newcommand{\SubAlign}[2]{\begin{subequations}\label{#1}\begin{align}#2\end{align}\end{subequations}}

\newcommand{\Figref}[1]{Fig.~\ref{#1}}
\newcommand{\Eqref}[1]{\eqref{#1}}

\newcommand{\bs}{\boldsymbol}

\newcommand{\groupU}[1]{\mathrm{U}(#1)}   		%Unitary group
\newcommand{\groupZ}[1]{\mathbb{Z}_{#1}} 		%Discrete group
\newcommand{\Exp}[1]{\text{e}^{#1}}

\renewcommand\Re{\mathrm{Re}}
\renewcommand\Im{\mathrm{Im}}
\newcommand{\Grad}{{\bs\nabla}}

\newcommand{\Curl}{{\bs\nabla}\times}

\newcommand{\F}{\mathcal{F}}
\newcommand{\Hc}[1]{\mathrm{H}_{c#1}} 
\newcommand{\D}{{\bs D}}
\newcommand{\A}{{\bs A}}
\newcommand{\B}{{\bs B}}
\newcommand{\J}{{\bs J}}
\newcommand{\Q}{\mathcal{Q}}

\begin{document}
%%%%%%%%%%%%%%%%%%%%%%%%%%%%%%%%%%%%%%%%%%%%%%%%%%%%%%%%%%%%%%%%%%%%%
%%%%%%%%%%%%%%%%%%%%%%%%%%%%%%%%%%%%%%%%%%%%%%%%%%%%%%%%%%%%%%%%%%%%%
%%%% Title informations and authors
\title{Properties of skyrmions and multi-quanta vortices 
	in chiral \texorpdfstring{$p$}{p}-wave superconductors}

\author{Julien~Garaud}
\email{garaud.phys@gmail.com}
\affiliation{Department of Theoretical Physics and  Center for Quantum Materials 
KTH-Royal Institute of Technology, Stockholm, SE-10691 Sweden}
\author{Egor~Babaev}%%\email{}
\affiliation{Department of Theoretical Physics  and Center for Quantum Materials 
KTH-Royal Institute of Technology, Stockholm, SE-10691 Sweden}

\date{\today}
%%%%%%%%%%%%%%%%%%%%%%%%%%%%%%%%%%%%%%%%%%%%%%%%%%%%%%%%%%%%%%%%%%%%%
%%%% The abstract
\begin{abstract}

Chiral $p$-wave superconducting state supports a rich spectrum of 
topological excitations different from those in conventional 
superconducting states. Besides domain walls separating different 
chiral states, chiral $p$-wave state supports both singular and 
coreless vortices also interpreted as skyrmions.
Here, we present a numerical study of the energetic properties of 
isolated singular and coreless vortex states as functions of 
anisotropy and magnetic field penetration length. In a given chiral 
state, single quantum vortices with opposite winding have different 
energies and thus only one kind is energetically favoured. We find 
that with the appropriate sign of the phase winding, two-quanta 
(coreless) vortices are always energetically preferred over two isolated 
single quanta (singular) vortices. We also report solutions carrying 
more flux quanta. However those are typically more energetically 
expensive/metastable as compared to those carrying two flux quanta.
\end{abstract}

\maketitle

Chiral $p$-wave superconducting state is an exotic state that, 
in addition to usual $\groupU{1}$ gauge symmetry, spontaneously 
breaks  time-reversal symmetry.
Higher broken symmetries there, implies a much richer spectrum 
of topological excitations as compared to conventional superconducting 
and superfluid states. Chiral $p$-wave pairing is realized in the 
A-phase of superfluid $^3$He, were variety of complex topological 
defects were investigated \cite{Volovik,Mermin.Ho:76,
Anderson.Toulouse:77,Mermin:79,Thuneberg:86,Walmsley.Cousins.ea:03,
Walmsley.White.ea:04,Walmsley.Golov:12}. 
In a superconducting $p$-wave state, due to the coupling to the 
vector potential, topological defects exhibit different properties.
This coupling affects their energy and determines their role in 
the magnetic properties of such superconductors. 
Layered perovskite superconductor Sr$_2$RuO$_4$ is a candidate 
material where various experimental evidences suggest possible 
realization of a $p$-wave superconducting state 
\cite{Maeno.Hashimoto.ea:94,Mackenzie.Maeno:03}. 
Similar models were also considered in connection to the 
superconducting state of heavy fermion compound UPt$_3$ 
\cite{Joynt.Taillefer:02,Joynt:91} (see e.g. \cite{Machida.Itoh.ea:12,
Tsutsumi.Machida.ea:12} for recent discussion of superconducting 
state in that material).

Spontaneous breaking of time-reversal symmetry for chiral 
$p$-wave state, implies the existence domain walls that separate 
regions with two different time-reversal symmetry broken (TRSB) 
ground-states. These domain walls, support spontaneous supercurrent 
that can generate magnetic fields \cite{Volovik.Gorkov:85,
Sigrist.Rice.ea:89,Matsumoto.Sigrist:99,Matsumoto.Sigrist:99a,
Bouhon.Sigrist:14,Scaffidi.Simon:15}. 
The domain walls in chiral $p$-wave superconductors could be 
created via Kibble-Zurek mechanism, and these properties can 
be used for their control \cite{Vadimov.Silaev:13}.
However, in Sr$_2$RuO$_4$ no indication of such a field was found 
in magnetic imaging microscopy experiments \cite{Kirtley.Kallin.ea:07,
Hicks.Kirtley.ea:10,Kallin.Berlinsky:09}.
Besides domain walls, chiral $p$-wave superconductors feature 
rich spectrum of topological defects including various vortices 
and skyrmions \cite{Knigavko.Rosenstein:99,Knigavko.Rosenstein.ea:99}. 
Related topological defects were also discussed in the context 
of heavy fermion superconductor UPt$_3$ \cite{Tokuyasu.Hess.ea:90}.

In zero field, both chiral (ground-)states are degenerate in
energy and this degeneracy is lifted by an externally applied 
magnetic field along the $c$-axis. For a given sign of the 
magnetic field parallel to the $c$-axis, only one of the chiral 
states is stable and the time-reversed state is energetically 
penalized. Likewise, vorticity of the superconducting condensates 
lifts the degeneracy between both chiral states. When the dominant 
component forms a vortex, it induces the time-reversed (subdominant) 
chiral component, in the vicinity of the core.
The winding of the induced component is not independent 
of that of the dominant component. It has a $4\pi$ winding 
of the relative phase between components, that follows from
the Cooper pairs having nonzero internal orbital momentum
\cite{Sauls:94}. Since the magnetic field lifts the 
degeneracy between chiralities, vortices and anti-vortices 
have different properties \cite{Sauls.Eschrig:09}.

It is experimentally seen that in an applied external field, 
Sr$_2$RuO$_4$ exhibits vortex lattices with square symmetry 
at high fields \cite{Riseman.Kealey.ea:98,Aegerter.Lloyd.ea:98,
Ray.Gibbs.ea:14}, and a transition to triangular lattice in 
lower fields \cite{Curran.Khotkevych.ea:11,Ray.Gibbs.ea:14}. 
Earlier theoretical calculations based on Ginzburg-Landau 
model for chiral $p$-wave superconductivity in Sr$_2$RuO$_4$ 
\cite{Agterberg:98,Agterberg:98a,Heeb.Agterberg:99}, are 
consistent with these observed transitions of the vortex lattice 
structure.

Besides single-quanta vortices, there also exists vortices 
carrying multiple quanta of the magnetic flux and that, as they 
are coreless, are essentially different from single-quanta 
vortices. 
For example as discussed in more details below, the component 
induced by a doubly quantized vortex in the dominant component 
has zero winding in subdominant one \cite{Sauls.Eschrig:09}. 
In this paper we demonstrate that the two-quanta (coreless) 
vortices, which can also be denoted as skyrmions, are energetically 
favoured as compared to (isolated) single-quanta vortices.
Earlier works in the context of UPt$_3$, even claim that lattices 
of similar two-quanta vortices may be energetically favoured as 
compared to those of single quanta \cite{Barash.MelNikov:90,
Melnikov:92}. The possible existence of lattices of different 
coreless vortices carrying single flux quantum in UPt$_3$ was 
also discussed recently \cite{Tsutsumi.Machida.ea:12}.   
It was also recently shown in the context of Sr$_2$RuO$_4$, based 
on solutions of microscopic Eilenberger equations, that lattices 
of two-quanta vortices are favoured for certain parameter sets
\cite{Ichioka.Machida.ea:12}. Yet, such lattices of two-quanta 
vortices were never observed in Sr$_2$RuO$_4$. This motivates 
this work to further investigate the thermodynamic stability of 
skyrmions for broad parameter range.

In a previous work \cite{Garaud.Babaev:12}, we reported isolated
skyrmion solutions in a model for chiral $p$-wave superconductor. 
For the studied case of one of the chiralities, skyrmions can be 
energetically favoured as compared to vortices (see also remark 
\footnote{
Note that the Ginzburg-Landau model which was used in Ref. 
\cite{Garaud.Babaev:12} had slightly different coefficients in 
the potential terms compared to the standard GL model which follows 
from the weak-coupling mean-field theory. In this paper we use the 
same model as in Ref.~\cite{Heeb.Agterberg:99}). 
}).
The skyrmions carrying two flux quanta are directly related to 
the two-quanta vortices mentioned above. However it was also 
demonstrated in Ref.~\cite{Garaud.Babaev:12} that there are 
(meta-)stable skyrmions carrying larger number of flux quanta. 
In this model the energy and structure of vortices and skyrmions 
depends on the chirality. Equivalently, for a given chiral state, 
vortex/skyrmion solutions are not the same as anti-vortex/anti-skyrmion. 
It thus calls for further investigation of vortex and skyrmion solutions 
(carrying two and more quanta), which we present below.

In the coordinate system where the crystal anisotropy axis is 
${\bf c}\parallel{\bf z}$, the order parameter of the $p_x+ip_y$ 
state is described by a complex two-dimensional vector 
${\bs\eta}=(\eta_x,\eta_y)$ \cite{Mackenzie.Maeno:03,Joynt.Taillefer:02,
Sigrist.Ueda:91}. Introducing the chiral order parameter components 
$\eta_\pm=\left(\eta_x\pm i\eta_y\right)/\sqrt{2}$, the dimensionless 
Ginzburg-Landau free energy reads as (see e.g. \cite{Agterberg:98,
Agterberg:98a,Heeb.Agterberg:99}): 
\SubAlign{Eq:FreeEnergy}{
&\mathcal{F}= 
	|\Curl\A|^2 + |\D\eta_+|^2 + |\D\eta_-|^2 
	\label{Eq:GradientEnergy1} \\
   &+(\nu+1)\Re\left[ (D_x\eta_+)^*D_x\eta_- - (D_y\eta_+)^*D_y\eta_- \right]	
    \label{Eq:GradientEnergy2} \\
   &+(\nu-1)\Im\left[ (D_x\eta_+)^*D_y\eta_- + (D_y\eta_+)^*D_x\eta_- \right]	
    \label{Eq:GradientEnergy3} \\
   &+2|\eta_+\eta_-|^2+\nu\Re\left(\eta_+^{*2}\eta_-^2\right)
   	+\sum_{a=\pm}-|\eta_a|^2+\frac{1}{2}|\eta_a|^4   
	\label{Eq:PotentialEnergy}	.
}
There we use dimensionless units were the free energy is normalized 
to the condensation energy, and the lengths are given in units of 
$\xi=\left(\alpha_0(T-T_c)\right)^{-1/2}$. The magnetic field $\B=\Curl\A$ 
is in units of $\sqrt{2}B_c=\Phi_0/(2\pi\lambda\xi)$. In these units, 
the gauge coupling $g$ that appears in the covariant derivative 
$\D=\Grad+ig\A$ is related to the Ginzburg-Landau parameter 
$g^{-1}=\lambda/\xi$. 
The free energy \Eqref{Eq:FreeEnergy} was derived from the weak coupling 
microscopic theory \cite{Agterberg:98,Agterberg:98a}. The anisotropy 
parameter $\nu$ determines the anisotropy in the $xy$-plane ($|\nu|<1$ 
for the energy to be positively defined). It measures the tetragonal 
distortions of the Fermi surface, which has cylindrical geometry for 
$\nu=0$, and is defined as $\nu=(\langle v_x^4\rangle-3\langle 
v_x^2v_y^2\rangle)/(\langle v_x^4\rangle+\langle v_x^2v_y^2\rangle)$ 
(where $\langle\cdot\rangle$ denote average over the Fermi surface). 
In the model Eq.~\Eqref{Eq:FreeEnergy}, the dependence on the third 
coordinate is not considered (i.e. assuming two-dimensional system or 
translational invariance along $z$-axis).

The ground-state that minimizes the potential terms in \Eqref{Eq:FreeEnergy} 
is degenerate and the solutions are $(\eta_+,\eta_-)=(1,0)$ and 
$(0,1)$. The theory \Eqref{Eq:FreeEnergy} is invariant under the 
(discrete) time-reversal operations $\mathcal{T}$, as 
$\{\eta_\pm,\B\}\xrightarrow{\mathcal{T}}\{\eta^*_\mp,-\B\}$. 
This invariance is spontaneously broken by the ground-state.
The spontaneous breakdown of the discrete time-reversal symmetry 
dictates that the theory allows domain wall solutions that 
interpolate between regions with different ground-states. 
Such domain walls, rather generically created during phase transition 
where the discrete symmetry is broken \cite{Vadimov.Silaev:13}, 
carry a magnetic field perpendicular to the $xy$-plane 
\cite{Matsumoto.Sigrist:99,Matsumoto.Sigrist:99a}. 
The discrete degeneracy of the ground state is lifted by the magnetic 
field. Thus, depending on the direction of the external field, only 
one state is stable. Likewise, the vorticity of the superconducting 
condensates lifts the degeneracy between chiral (ground-)states.

As the components $\eta_+$ and $\eta_-$ behave differently for different 
sign of the winding, a complete study requires to consider both situations 
of counter-clockwise (positive) and clockwise (negative) vorticities. 
Note that this is equivalent to considering only positive vorticity but 
for both chiral states. 
For example, the configuration with a winding $n_+=+1$ on the ground-state 
$(\eta_+,\eta_-)=(1,0)$ can be obtained by applying the time-reversal 
operation $\mathcal{T}$ on the configuration whose ground state is 
$(\eta_+,\eta_-)=(0,1)$ with the winding $n_-=-1$.
In the following, we choose to fix the dominant component of the order 
parameter to be $\eta_-$ [i.e. the ground state is 
$(\eta_+,\eta_-)=(0,1)$] and thus need to investigate both positive 
and negative vorticities.

The asymptotic vorticity of the dominant component $\eta_-$ 
determines the sign of $B_z$, as well as the vorticity of 
the subdominant component $\eta_+$ \cite{Agterberg:98a}, 
according to:
\Equation{Eq:vorticity}{
\eta_-\propto\Exp{in_-\theta}\,,~
\eta_+\propto\Exp{in_+\theta} 
~~~\text{and}~~~n_+=n_-+2\,,
}
where $\theta$ is the polar angle. 
The relative phase between $\eta_+$ and $\eta_-$ \Eqref{Eq:vorticity},
follows from the internal orbital momentum of Cooper pairs. In the 
Ginzburg-Landau model \Eqref{Eq:FreeEnergy}, this is the structure 
of mixed gradient that constraints the relative phase. Note that 
since the subdominant component $\eta_+$ vanishes asymptotically 
[i.e. it recovers its ground state value $\eta_+=0$], the winding 
$n_+$ can be located only in a close vicinity of the vortex core. 
Note also that the winding of the subdominant component does not 
affect the overall flux quantization, because the density of that 
component vanishes away from the vortex. From \Eqref{Eq:vorticity} 
it is rather straightforward to see that the vortex with the 
vorticity $(n_+,n_-)=(+3,+1)$ and the (anti-)vortex with 
$(n_+,n_-)=(+1,-1)$ have different core structures and thus 
different energy. 

%%%%%%%%%%%%%%%%%%%%%%%%%%%%%%%%%%%%%%%%%%%%%%%%%%%%%%%%%%%%%
\begin{figure*}[!htb]
\hbox to \linewidth{ \hss
\includegraphics[width=\linewidth]{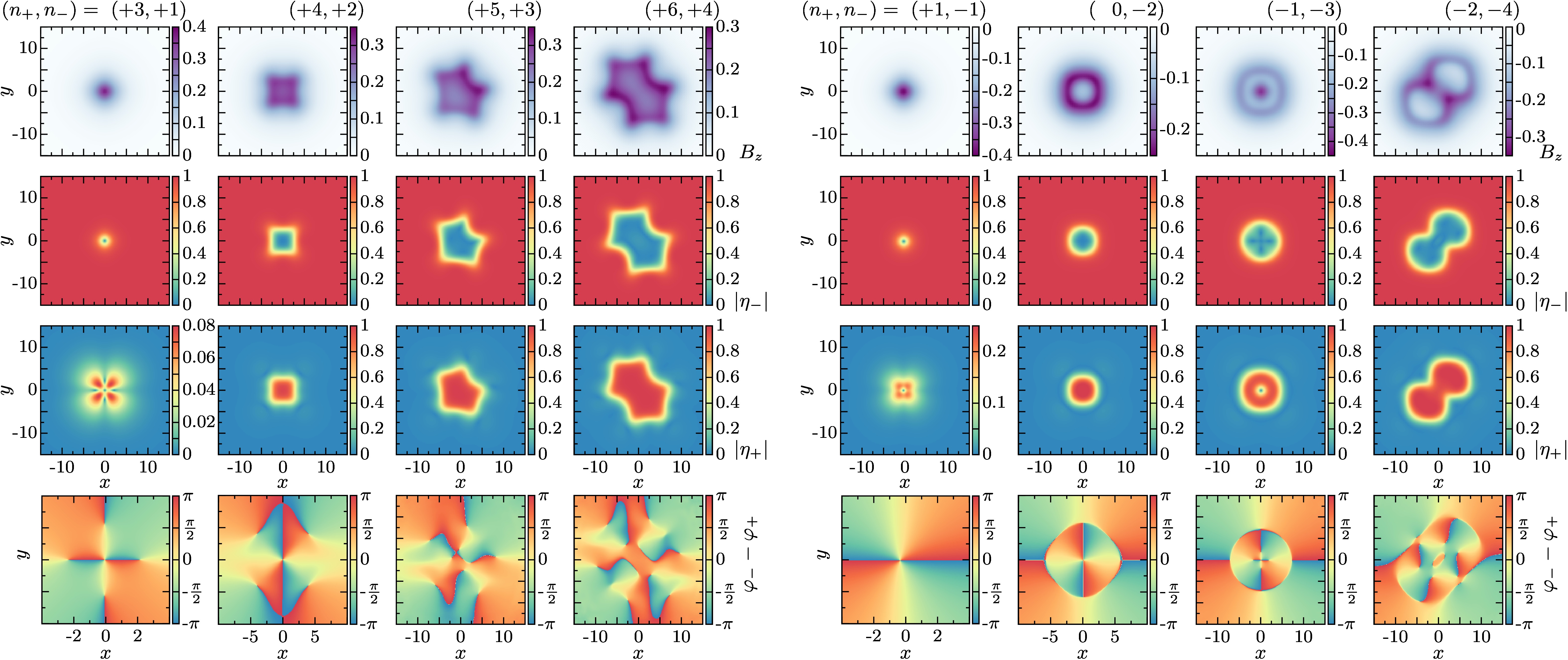}
\hss}
\caption{
(Color online) -- 
Vortex states for the windings $(n_+,n_-)$ of the components 
$\eta_+$ and $\eta_-$. Note that the winding of the dominant 
component (here $\eta_-$), specifies the flux carried by the 
vortex configuration. The parameters of the Ginzburg-Landau 
functional are $g=0.3$ and $\nu=0.2$. 
The first line shows the magnetic field $\B$, while second and 
third line respectively display $|\eta_-|$ and $|\eta_+|$. The 
fourth line shows the relative phase $\varphi_--\varphi_+$ 
between $\eta_+$ and $\eta_-$. Winding of the relative phase 
indicates the position of the cores of $\eta_+$ and $\eta_-$. 
The first block shows vortex solutions carrying one to four 
flux quanta with $B_z>0$, while the second block shows the 
corresponding vortices with $B_z<0$.
}
\label{Fig:Vortices}
\end{figure*}
%%%%%%%%%%%%%%%%%%%%%%%%%%%%%%%%%%%%%%%%%%%%%%%%%%%%%%%%%%%%% 

In order to investigate the energetic properties of vortex
matter, the fields $\eta_\pm$ and $\A$ are discretized using 
a finite-element framework \cite{Hecht:12} and the free energy 
\Eqref{Eq:FreeEnergy} is minimized using an nonlinear conjugate 
gradient algorithm (see the appendix for details).
In simulations of chiral $p$-wave superconductors on a finite 
domain, a special attention is required for boundary conditions 
in order to yield edge currents (see for example discussions in 
\cite{Sigrist.Ueda:91,Sauls:11,Bouhon.Sigrist:14}).
Here, we are interested in the intrinsic energetic properties of 
isolated defects. Thus vortex configuration is created by an 
initial guess and placed them at the center of a large domain, 
with open boundary conditions, letting the fields freely reach 
the ground-state. By choosing a sufficiently large domain, this 
ensures that within numerical accuracy  vortices will not interact 
with boundaries and thus we are able to probe their intrinsic 
structure and energy properties, without effects of boundary 
conditions.
As it specifies the topological sector, a starting configuration 
with a given winding $n_-$ of the dominant component $\eta_-$ leads, 
after convergence of the algorithm, to a configuration that behaves 
as expected from \Eqref{Eq:vorticity}.
We systematically construct vortex solutions carrying one to  
four flux quanta for parameter space defined by wide range of 
values of the $g$ and $\nu$. 
\Figref{Fig:Vortices} shows typical vortex solutions with 
different vorticities. Along this paper we also refer to vortices 
carrying multiple flux quanta, as \emph{skyrmions}. The reason for 
that terminology is that they have additional topological properties, 
as compared to single quanta vortices. This is explained in more 
details by the end of the paper.

%%%%%%%%%%%%%%%%%%%%%%%%%%%%%%%%%%%%%%%%%%%%%%%%%%%%%%%%%%%%%
\begin{figure*}[!htb]
\hbox to \linewidth{ \hss
\includegraphics[width=0.8\linewidth]{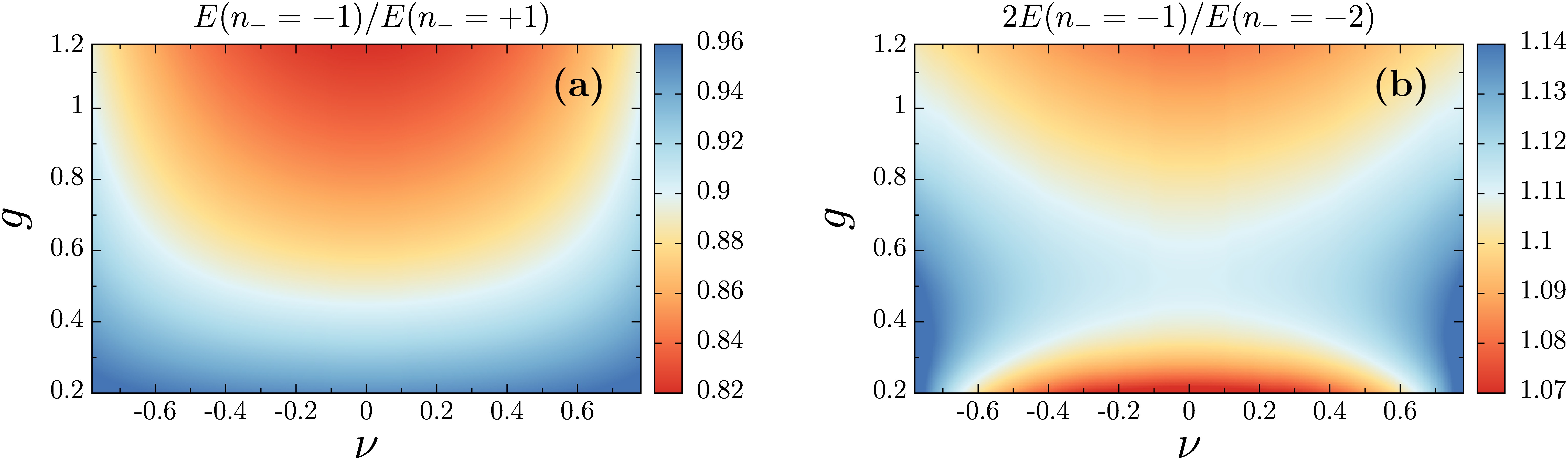}
\hss}
\caption{
(Color online) -- 
Left panel shows the ratio of the energies $E(n_-=-1)$ of $n_-=-1$ 
single quanta vortices and $E(n_-=+1)$ of $n_-=+1$ vortices, as 
functions of the anisotropy parameter $\nu$ and of the gauge coupling 
$g$. This ratio is always smaller than $1$ implying that 
$(n_+,n_-)=(+1,-1)$ are always less energetic than $(n_+,n_-)=(+3,+1)$. 
The right panel displays the ratio of the energies $2E(n_-=-1)$ of 
two $n_-=-1$ single-quanta vortices and $E(n_-=-2)$ of one $n_-=-2$ 
vortex, as functions of the anisotropy parameter $\nu$ and of the 
gauge coupling $g$. This ratio is always larger than $1$ implying 
that two-quanta vortices are less energetic than two isolated 
single-quanta vortices.
}
\label{Fig:DiagramMinus}
\end{figure*}
%%%%%%%%%%%%%%%%%%%%%%%%%%%%%%%%%%%%%%%%%%%%%%%%%%%%%%%%%%%%%

The first and second blocks in \Figref{Fig:Vortices} respectively show 
vortex solutions with $B_z>0$ and $B_z<0$. Vortices carrying from one 
to four flux quanta are displayed within each block. As expected from 
Eq.~\Eqref{Eq:vorticity}, single winding of the dominant component 
induces core structure of the subdominant component with different 
winding depending on that of $\eta_-$ (see the first column of each block). 
It is instructive to consider the last row in \Figref{Fig:Vortices}, 
that displays the relative phase between $\eta_-$ and $\eta_+$. 
In agreement with \Eqref{Eq:vorticity}, asymptotically the relative 
phase $\varphi_--\varphi_+=-2\theta$, reflecting the orbital angular 
momentum difference between $\eta_-$ and $\eta_+$. Moreover, the 
relative phase also indicates the position of the singularities of 
the components of the order parameter. Remarkably single quanta 
vortices are singular defects, since singularities in both 
components overlap (and thus $\eta_+=\eta_-=0$). On the other hand, 
since both components never simultaneously vanish, two-quanta vortices 
are coreless defects. Interestingly the $n_-=-2$ configuration 
features a $\pi$ jump of the relative phase when going outward 
from the vortex core. Inside the vortex core the time-reversed 
chiral state $(\eta_+,\eta_-)=(1,0)$ is induced, while the 
$(\eta_+,\eta_-)=(0,1)$ state is recovered asymptotically. 
The two quanta vortices thus feature a domain wall when going 
away from the center. The $\pi$ jump of the relative phase
for the $n_-=-2$ is located at this domain wall.

Like in conventional superconductors, the magnetic field for 
single quanta vortices is localized at the vortex core and 
screened at length scales determined by the penetration depth 
$\lambda$. Interestingly, the magnetic field for two-quanta 
vortices, and especially for $n_-=-2$, is not homogeneously 
distributed in the core. Rather it is localized at a given 
distance from the center and spread along the ring of the 
domain wall. Note that similar vortex configurations were also 
found to exist in the context of two-component model with 
$\groupU{1}\times\groupU{1}\times\groupZ{2}$ symmetry 
\cite{Garaud.Babaev:14b}. The ring-like distribution of the 
magnetic field for the two-quantum vortex can be understood 
as follows: $\B$ outside the vortex is screened by the 
(partial) currents in $\eta_-$ that run counter-clockwise, 
while inside the vortex currents in $\eta_+$ are responsible 
for the screening. 
Since $\eta_+$ vanishes away from the vortex core, it cannot 
contribute to the screening asymptotically. Conversely, $\eta_-$ 
vanishes at the vortex core and this is the induced subdominant 
component $\eta_+$ that screens $\B$  close to the center of 
the vortex core. The reason it can contribute to screening 
(inside the vortex) without having vorticity on its own, is only 
due to supercurrent produced by the vector potential (like the 
Meissner currents on the boundary of ordinary superconductors).
Since those currents circulate clockwise, they compensate with 
the currents in $\eta_-$ so that at a certain distance (at the 
domain wall) there is no screening current. The magnetic field 
is thus localized at the domain wall. 
Although the core structure of single-quanta vortices are 
different depending on the sign of $n_-$, their profile 
of the magnetic field looks quite similar. When considering 
vortices with $|n_-|>1$, both the core structure and the 
magnetic field profile are strikingly different and the 
skyrmions with negative $n_-$ do not resemble those with 
positive $n_-$.
Apart from the $n_-=-2,-3$ skyrmions, the configurations 
that carry multiple flux quanta are far from being axially 
symmetric. 
Note that the $n_-=-4$ skyrmions resembles as some kind 
of bound state of two $n_-=-2$ skyrmions. As we will 
discuss below, this makes their decay into two $n_-=-2$ 
vortices rather easy.

%%%%%%%%%%%%%%%%%%%%%%%%%%%%%%%%%%%%%%%%%%%%%%%%%%%%%%%%%%%%%
\begin{figure*}[!htb]
\hbox to \linewidth{ \hss
\includegraphics[width=\linewidth]{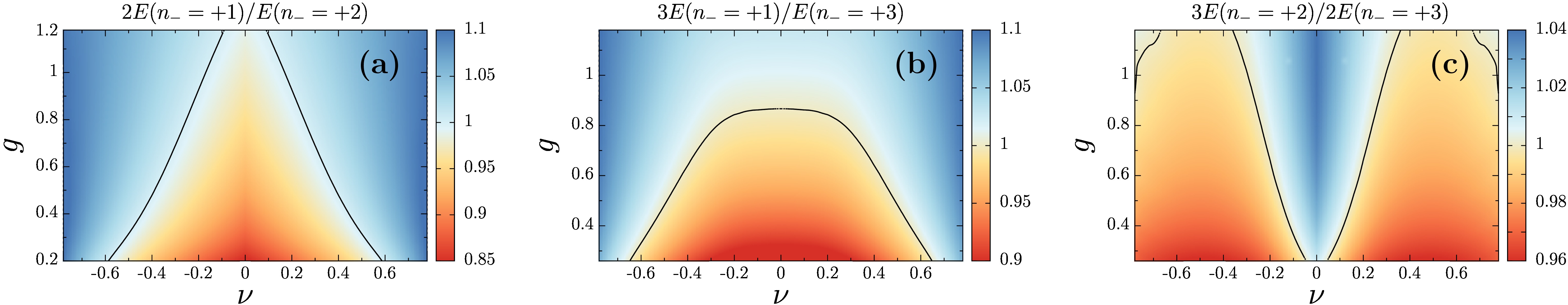}
\hss}
\caption{
(Color online) -- 
Ratio of the energies of multiple quanta vortices with $n_->0$
compared to isolated vortices carrying smaller number of flux 
quanta, as functions of the anisotropy parameter $\nu$ and of 
the gauge coupling $g$. The solid line separates the regions 
where isolated vortices with smaller flux are preferred over 
single vortex carrying more flux quanta.
Panel (a) and (b) respectively show the energetic behaviour 
of double (resp. triple) quantum vortex compared to isolated 
vortices with single flux quantum. Below the solid line, that 
is for more isotropic (small $|\nu|$) or more type-2 (small $g$), 
isolated single quanta vortices are energetically favoured. 
Panel (c) shows the energy ratio of two (isolated) triple-quanta 
vortices compared to three double quanta.
This indicates subtle sublinear energy scaling with the number 
of flux quanta.
}
\label{Fig:DiagramPlus}
\end{figure*}
%%%%%%%%%%%%%%%%%%%%%%%%%%%%%%%%%%%%%%%%%%%%%%%%%%%%%%%%%%%%%

Since the core structure is different, it is quite natural to 
expect that, unlike in conventional superconductors, vortices 
with opposite winding (and thus opposite directions of the 
magnetic field) are non-degenerate in energy. The 
$(n_+,n_-)=(+3,+1)$ vortex has more total vorticity than the 
$(n_+,n_-)=(+1,-1)$. Thus one could naively expect that the 
$n_-=-1$ vortices would be favoured as compared to $n_-=+1$. 
We systematically compared the energies of both single-quanta 
vortices for all values of the anisotropy parameter $\nu$ and 
of the gauge coupling $g$.
The diagram in \Figref{Fig:DiagramMinus}(a), shows that the 
ratio of the energies of the single quanta vortices with $n_-=-1$ 
and $n_-=+1$, is always less than one. This implies that vortices 
$n_-=-1$ are always energetically favoured, as compared to those 
with $n_-=+1$. The first critical field of a vortex carrying a 
flux $\Phi$ is $\Hc{1}=E/2\Phi$, where $E$ is its energy. As a 
result, \Figref{Fig:DiagramMinus}(a) also implies that the $n_-=-1$ 
vortices also have lower first critical field $\Hc{1}$ in agreement 
with Refs.~\cite{Heeb.Agterberg:99,Ichioka.Matsunaga.ea:05}. 
Although both $n_-=\pm 1$ are perturbatively stable (i.e. they 
are minima of $\F$), only $n_-=-1$ is absolutely stable.

Note that the naive estimates based on counting the total vorticity 
provide the correct picture that $(n_+,n_-)=(+1,-1)$ vortices 
are less energetic than $(n_+,n_-)=(+3,+1)$ ones. It thus makes 
sense to apply the same arguments to configurations carrying more 
than one flux quantum. In the sector with negative $n_-$, there 
are two possibilities to make a configuration that carries two 
flux quanta. Either to create two isolated $(n_+,n_-)=(+1,-1)$ 
vortices carrying one flux quantum each or to create one 
$(n_+,n_-)=(0,-2)$ two-quantum vortex. 
It turns out that a two-quantum vortex with smaller number of 
singularities is favoured as compared to two isolated single-quanta 
vortices. \Figref{Fig:DiagramMinus}(b) displays the ratio of 
the energies of two (isolated) $n_-=-1$ vortices and one $n_-=-2$ 
vortex. This ratio is always larger than one, thus implying that 
two-quanta vortices are energetically favoured as compared to two 
isolated single-quanta vortices.
Note that the quantity displayed in \Figref{Fig:DiagramMinus}(b), 
is actually also the ratio of first critical fields associated with 
single and double quanta vortices $\Hc{1}(n_-=-1)/\Hc{1}(n_-=-2)$.
Note also that smaller $H_{c1}$ for a  higher-flux vortex does not 
necessarily imply that such vortices will nucleate first in low 
magnetic field. That is, due to higher winding they carry larger 
magnetic flux and thus can have a higher potential barrier to enter 
the sample (compared with the discussion of  Bean-Livingston barrier 
in single component superconductors \cite{Bean.Livingston:64}).
The vortices $(n_-=-1)$ and $(n_-=-2)$ should interact differently 
with the Meissner currents and image charges, and thus even if the 
$(n_-=-2)$ vortices have lower $\Hc{1}$, the interaction with the 
boundary may instead favour the entry of the vortices with $(n_-=-1)$.

We also calculated the energy diagram similar to that in 
\Figref{Fig:DiagramMinus}(b), but for vortices carrying three 
flux quanta $n_-=-3$ (data not shown). We found that unlike for 
$n_-=-2$, the $n_-=-3$ are not always stable. That is, in some 
regions of the parameter space the $n_-=-3$ is found, but in 
some other regions it decays into one single-quantum plus one 
double-quantum vortex. We find that for $n_-<0$, the $n_-=-2$ 
skyrmions are all energetically favoured. This behaviour can 
already be anticipated from the last column in \Figref{Fig:Vortices} 
where the four quanta $n_-=-4$ skyrmion seems to be a bound state 
of two $n_-=-2$ vortices. As the skyrmions with $n_-<-2$ are more 
energetic than those with $n_-=-2$, one can easily see that the 
$n_-=-4$ configuration can decay into two $n_-=-2$ vortices and 
thus reduce its total energy.
We find that in the regions where the $n_-=-3$ vortices exist, 
they are more energetic than three isolated single quanta vortices 
(or one single plus one two-quantum vortex).

Although being energetically unfavoured, it is still instructive 
to consider the properties of vortices carrying multiple flux quanta, 
with $n_->0$. Diagrams in \Figref{Fig:DiagramPlus} show the ratio 
of the energies of multiple quanta vortices with $n_->0$, compared 
to that of isolated vortices carrying smaller flux. 
The situation for $n_->0$ is actually very different from that with 
$n_-<0$. Panels (a) and (b) in \Figref{Fig:DiagramPlus} display
the ratio of the energies of isolated single-quanta vortices with 
that of vortices carrying two and three flux quanta. Depending 
on the anisotropy parameter $\nu$ and on the gauge coupling $g$
this ratio can either be smaller or larger than one and the solid 
lines on the diagram show when these are degenerate in energy.
Below the solid line, isolated single-quanta vortices are 
energetically favoured as compared to multi-quanta vortices.
Above this line, these are the vortices carrying two or three 
flux quanta which are favoured. Thus tetragonal distortions 
of the Fermi surface (i.e. larger $|\nu|$) tend to favour $n_-=+2$ 
(and to a lesser extend $n_-=+3$), as compared to isolated 
$n_-=+1$ vortices. 
Note that the solid lines in panel (a) and (b) do not coincide. 
The panel (c) shows the comparison between three isolated double 
quanta and two isolated triple quanta vortices. Here again, depending 
on $\nu$ and $g$, either can be preferred. This suggests complicated 
sublinear scaling of the energy with the number of flux quanta.

%%%%%%%%%%%%%%%%%%%%%%%%%%%%%%%%%%%%%%%%%%%%%%%%%%%%%%%%%%%%%
\begin{figure*}[!htb]
\hbox to \linewidth{ \hss
\includegraphics[width=\linewidth]{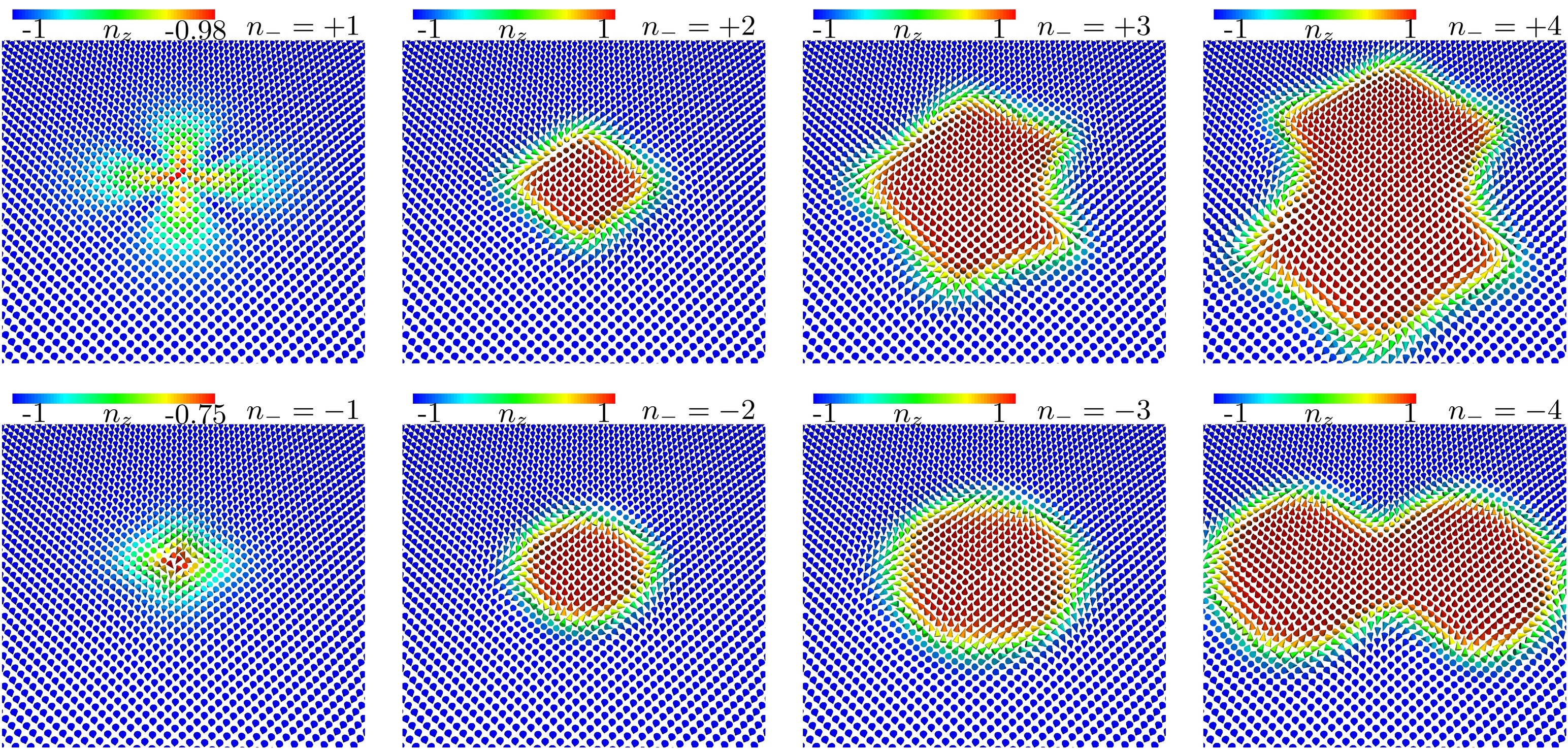}
\hss}
\caption{
(Color online) -- 
Pseudo-spin texture $\bs n$ defined as the projection of superconducting 
degrees of freedom onto spin-$1/2$ Pauli matrices. Different panels 
correspond to the solutions with different vorticity displayed in 
\Figref{Fig:Vortices}. First line shows vortices with $B_z>0$ and the 
second line is for $B_z<0$. 
The multi-quanta skyrmions are characterized by the topological 
charge \Eqref{Eq:Charge} $\Q=n_-$. Single quanta vortices on the other 
hand, do not cover the whole sphere (i.e $n_z\leq n_z^\mathrm{max}<1$) 
thus they have vanishing charge $\Q=0$.
}
\label{Fig:Texture}
\end{figure*}
%%%%%%%%%%%%%%%%%%%%%%%%%%%%%%%%%%%%%%%%%%%%%%%%%%%%%%%%%%%%%

The coreless nature of the two-quanta vortices implies that these have 
additional topological properties that are absent for single-quanta 
vortices. If the order parameter $\bs{\tilde{\eta}}=(\eta_+,\eta_-)$ 
does not vanish ($\bs{\tilde{\eta}}\neq0$), a pseudo-spin (unit) vector 
$\bs n$ can be defined as the projection of the order parameter on 
spin-1/2 Pauli matrices ${\bs\sigma}$: 
$n_i=\bs{\tilde{\eta}}^\dagger\sigma_i\bs{\tilde{\eta}}/
\bs{\tilde{\eta}}^\dagger\bs{\tilde{\eta}}$ 
(see detailed discussion of the pseudo-spin formalism for 
multi-component GL models in \cite{Babaev.Faddeev.ea:02}).
\Figref{Fig:Texture} shows pseudo-spin texture for vortex solutions 
corresponding to those displayed in \Figref{Fig:Vortices}. Note 
that $\bs n$ is ill-defined for singular vortices, since there 
$\bs{\tilde{\eta}}=0$ at the core (i.e. singularities in both 
components overlap). Coreless vortices on the other hand 
have well defined pseudo-spin projection which is a map 
$\bs n:\mathbbm{R}^2\to S^2$. Since at spatial infinity, 
$\bs n=(0,0,-1)$, the plane $\mathbbm{R}^2$ can be compactified 
to $S^2$ so that the pseudo-spin becomes a map $\bs n:S^2\to S^2$. 
The homotopy invariants $\pi_2(S^2)\in\mathbbm{Z}$ associated with 
such maps defines the integer-valued topological charge 
\Equation{Eq:Charge}{
\Q=\frac{1}{4\pi}\int_{\mathbbm{R}^2}
	{\bs n}\cdot\partial_x{\bs n}\times\partial_y{\bs n}\,\,dx\,dy\,,
}
which can be used to classify various field configurations. 
Heuristically, $\Q$ counts the number of times that the target 
sphere $S^2$ is wrapped while covering the $xy$-plane. Singular 
configurations for which the pseudo-spin is not everywhere 
well-defined, have $\Q=0$. Non-singular solutions on the other hand, 
and in particular coreless vortices, have $\Q=g\Phi/2\pi\in \mathbbm{Z}$ 
(where $\Phi$ is the flux). For example the two-quanta vortices, 
which are coreless, are characterized by $\Q=2$.
The fact that $\Q=0$ for singular vortices can easily be seen from 
the plot of the pseudo-spin texture \Figref{Fig:Texture}. There 
$\bs n$ never reaches the north pole and thus do not fully cover 
the unit sphere.

Here, we reported a large-scale numerical investigation of the energy 
properties of isolated single and multiple quanta vortices/skyrmions 
in a Ginzburg-Landau model of chiral $p$-wave superconducting state. 
As pointed out previously, for a given ground-state chirality, vortices 
and anti-vortices are inequivalent. Thus we performed study for both 
orientations of the winding. 
The vortices with winding $n_-=-1$ in the dominant component are always 
preferred to those with winding $n_-=+1$. We also found that vortices 
carrying two flux quanta with $n_-=-2$ are always energetically favoured 
as compared to two isolated single-quanta vortices. Vortices with higher 
flux and negative $n_-$, on the other hand, are either unstable or 
have higher energies per flux quantum.
We also reported the structural and energetic properties of 
(meta-)stable skyrmions with various topological charge (i.e. for 
$n_->+1$). The calculations show complicated sublinear scaling of 
the energy with the number of flux quanta that qualitatively agrees 
with previous works for a smaller parameter set in a related model 
\cite{Garaud.Babaev:12}. 
Due to their very characteristic profiles of the magnetic field, their 
experimental observation, in e.g. scanning Hall and scanning SQUID 
experiments would provide a strong evidence of chiral $p$-wave 
superconductivity in the candidate materials described by the model 
\Eqref{Eq:FreeEnergy}. 
Note however that various aspects of microscopic physics may alter 
the form of the Ginzburg-Landau model \Eqref{Eq:FreeEnergy}, and in 
particular the balance between the different coefficients entering the
free energy. This is currently a subject of ongoing studies, in connection 
with Sr$_2$RuO$_4$  (see e.g. \cite{Bouhon.Sigrist:14,Scaffidi.Simon:15}).
Note added: after the completion of this work a study appeared 
reporting stable skyrmions as well as vortices, in this model,
affected by mesoscopic effects in small samples
\cite{FernandezBecerra.Sardella.ea:16}. 

\begin{acknowledgments}
We acknowledge useful discussions with Troels Bojesen, Adrien Bouhon, 
Mihail Silaev and Asle Sudb{\o} .
The work was supported by the Swedish Research Council grants 
642-2013-7837,  325-2009-7664. 
The computations were performed on resources provided by the 
Swedish National Infrastructure for Computing (SNIC) at National 
Supercomputer Center at Link\"oping, Sweden.
\end{acknowledgments}

%%%%%%%%%%%%%%%%%%%%%%%%%%%%%%%%%%%%%%%%%%%%%%%%%%%%%%%%%%%%%%%%%%%%%
\appendix
\section{Numerical Methods}

In this work we used the dimensionless two-component Ginzburg-Landau 
theory Eq.\Eqref{Eq:FreeEnergy} that was previously microscopically 
derived in the weak coupling limit (see for example 
Refs.~\cite{Agterberg:98,Agterberg:98a}).
In this work, we focus on the properties of vortex solutions in the 
$xy$-plane and neglect the dependence over the third coordinate $z$. 
This means that our solutions are either purely two dimensional, 
or describe bulk configuration, assuming translation invariance 
along $z$-axis (and thus neglecting possible surface effects).

For the numerical investigation, the two-dimensional problem 
\Eqref{Eq:FreeEnergy} is defined on a bounded domain 
$\Omega\subset\mathbbm{R}^2$. 
The boundary conditions for chiral $p$-wave superconductors can be 
very involved. Namely, in order to simulate chiral $p$-wave 
superconductors on a finite domain, a special attention has to be 
paid to boundary conditions to take into account edge currents 
properties. However, we are interested here in the intrinsic 
energetic properties of isolated defects. Thus we consider 
isolated vortices in large grids (such that there are no interactions 
with boundaries) and let the fields freely recover the ground-state. 
As a result, we probe the intrinsic structure and energy properties 
of vortices without any deformation originating from boundary behaviour.
The simulation is run for a zero applied field (so that there are 
no Meissner currents), and the flux carrying solution is generated 
by a starting condition with a given winding of the dominant 
component. 
Because it enjoys topological protection, the (dominant) component 
cannot unwind by means of continuous transformations and thus 
topological properties (winding of the dominant component) are 
preserved by an energy minimization algorithm. Note that as simulations 
are run on a large but finite domain, there is still a possibility 
to change the topological sector, by moving the vortex across the 
boundary. This is possible, because without external fields there are 
no Meissner currents to prevent escape of a vortex. Note however 
that as we choose to work with large grids, the vortices in practice 
do not interact with boundaries, and thus they do not escape from 
the domain. The advantage of this choice is that it is guaranteed 
that obtained solutions are not affected by boundaries and that the 
calculated energies are those of isolated defects. The configurations 
displayed in the paper are close-up views of these defects.

For the actual numerical computation, the variational problem of 
minimizing the free energy is defined using a finite element 
formulation provided by the {\tt Freefem++} library \cite{Hecht:12}.
Discretization within finite element 
formulation is done via a (homogeneous) triangulation over $\Omega$, 
based on Delaunay-Voronoi algorithm. Functions are decomposed 
over a continuous piecewise quadratic basis on each triangle. 
The accuracy of such method is controlled through the number of 
triangles, (we typically used $3\sim6\times10^4$), the order of 
expansion of the basis on each triangle (2nd order polynomial basis 
on each triangle), and also the order of the quadrature formula for 
the integral on the triangles. 
A nonlinear conjugate gradient algorithm is used to solve the variational 
nonlinear problem (i.e. to find the minima of $\F$). The algorithm is 
iterated until relative variation of the norm of the gradient of 
the functional $\F$ with respect to all degrees of freedom is less 
than $10^{-8}$ (we verified that for this value, the configuration 
does not evolve and the energy remains constant).

For the minimization procedure to lead to a configuration that 
has the expected topological properties, the starting field 
configuration should exhibit itself those desired topological 
properties. Although strictly speaking there is no infinite 
energy barrier between different topological sectors in finite 
domains, the barrier   is finite but large enough to 
prevent any unwinding. Thus typically gradient minimization 
converges to the configuration that has the topological properties 
of the starting guess. In order to have efficient numerics, it is 
also important that the starting field configuration is the closest 
as possible to the minimal energy configuration. 
The initial field configuration carrying $N_v$ flux quanta is 
prepared by using an ansatz which imposes phase winding of the 
dominant component ($\eta_-=|\eta_-|\Exp{i\varphi_-}$) around 
a given point $(x_k,y_k)$:
\Align{Eq:Guess}{
|\eta_-| &=\prod_{k=1}^{N_v} 
\sqrt{\frac{1}{2} \left( 1+\tanh\left(\frac{4}{\xi_a}({\cal R}_k(x,y)-\xi_a) 
\right)\right)}\, \\
\varphi_-&=\sum_{k=1}^{N_V}
      \tan^{-1}\left(\frac{y-y_k}{x-x_k}\right) 
~\text{and}~|\eta_+| =1-|\eta_-|\, ,
}
where ${\cal R}_k(x,y)=\sqrt{(x-x_k)^2+(y-y_k)^2}$ and $\xi_a$ 
parametrizes the core size. The parametrization of $\eta_+$, with 
nonzero density in the core enhances the convergence to form coreless 
defects. Finally, the starting configuration for the vector potential 
of the magnetic field $\A$, is determined by solving Amp\`ere's law 
equation $\Curl\B+\J=0$, for the supercurrent $\J=\delta\F/\delta\A$ 
specified by the superconducting condensates given by \Eqref{Eq:Guess}. 
Being an equation linear in $\A$, this operation is rapidly solved. 
Once the starting configuration is constructed, all degrees of 
freedom are relaxed simultaneously.

%%%%%%%%%%%%%%%%%%%%%%%%%%%%%%%%%%%%%%%%%%%%%%%%%%%%%%%%%%%%%%%%%%%%%
%%%%%%%%%%%%%%%%%%%%%%%%%%%%%%%%%%%%%%%%%%%%%%%%%%%%%%%%%%%%%%%%%%%%%
%%\bibliographystyle{apsrev4-1}
%\bibliography{../SRO-skyrmions-2015}

\end{document}